\documentclass[11pt]{article}
\usepackage{amssymb,amsmath,bm,graphicx}
\usepackage{mathrsfs}
\usepackage{constants}
\usepackage{hyperref}
\usepackage{enumitem}
\usepackage{graphics,lipsum}
\usepackage{float}
\usepackage{amssymb,amsmath,bm,graphicx}
\usepackage{natbib}
\usepackage{constants}
\usepackage{url}
\usepackage{color}

\date{}
\begin{document}
\newcommand{\bea}{\begin{eqnarray}}
\newcommand{\ena}{\end{eqnarray}}
\newcommand{\beas}{\begin{eqnarray*}}
\newcommand{\enas}{\end{eqnarray*}}
\newcommand{\beq}{\begin{equation}}
\newcommand{\enq}{\end{equation}}
\def\qed{\hfill \mbox{\rule{0.5em}{0.5em}}}
\newcommand{\bbox}{\hfill $\Box$}
\newcommand{\ignore}[1]{}
\newcommand{\ignorex}[1]{#1}
\newcommand{\wtilde}[1]{\widetilde{#1}}
\newcommand{\qmq}[1]{\quad\mbox{#1}\quad}
\newcommand{\qm}[1]{\quad\mbox{#1}}
\newcommand{\nn}{\nonumber}
\newcommand{\Bvert}{\left\vert\vphantom{\frac{1}{1}}\right.}
\newcommand{\To}{\rightarrow}
\newcommand{\E}{\mathbb{E}}
\newcommand{\Var}{\mathrm{Var}}
\newcommand{\Cov}{\mathrm{Cov}}
\newcommand{\Corr}{\mathrm{Corr}}
\newcommand{\dist}{\mathrm{dist}}
\newcommand{\diam}{\mathrm{diam}}
\makeatletter
\newsavebox\myboxA
\newsavebox\myboxB
\newlength\mylenA
\newcommand*\xoverline[2][0.70]{%
    \sbox{\myboxA}{$\m@th#2$}%
    \setbox\myboxB\null
    \ht\myboxB=\ht\myboxA%
    \dp\myboxB=\dp\myboxA%
    \wd\myboxB=#1\wd\myboxA
    \sbox\myboxB{$\m@th\overline{\copy\myboxB}$}
    \setlength\mylenA{\the\wd\myboxA}
    \addtolength\mylenA{-\the\wd\myboxB}%
    \ifdim\wd\myboxB<\wd\myboxA%
       \rlap{\hskip 0.5\mylenA\usebox\myboxB}{\usebox\myboxA}%
    \else
        \hskip -0.5\mylenA\rlap{\usebox\myboxA}{\hskip 0.5\mylenA\usebox\myboxB}%
    \fi}
\makeatother

\newtheorem{theorem}{Theorem}[section]
\newtheorem{corollary}[theorem]{Corollary}
\newtheorem{conjecture}[theorem]{Conjecture}
\newtheorem{proposition}[theorem]{Proposition}
\newtheorem{lemma}[theorem]{Lemma}
\newtheorem{definition}[theorem]{Definition}
\newtheorem{example}[theorem]{Example}
\newtheorem{remark}[theorem]{Remark}
\newtheorem{case}{Case}[section]
\newtheorem{condition}{Condition}[section]
\newcommand{\proof}{\noindent {\it Proof:} }

\title{{\bf\Large Do elderly want to work? Modeling elderly's decision to fight aging Thailand}}
\author{Krittiya Kantachote\thanks{Department of Sociology, Srinakharinwirot University, Bangkok, Thailand. Email: krittiyak@g.swu.ac.th} and Nathakhun Wiroonsri \thanks{Mathematics and Statistics with Applications Research Group. Department of Mathematics, King Mongkut's University of Technology Thonburi, Bangkok 10140, Thailand. Email: nathakhun.wir@kmutt.ac.th }  \\ Srinakharinwirot University and King Mongkut's University of Technology Thonburi}

\footnotetext{ Elderly retirement, lasso logistic regression, machine learning, random forest, Thailand}

\maketitle

\begin{abstract} 
Thailand has entered into an aging society since the year 2000. Using the 2017 Survey of the Older Persons in Thailand collected by Thailand National Statistical Office, this study uses cross tabulation, random forest with variable importance measure and lasso logistic regression to examine factors that have effects on the elderly's decision to remain in the labor market after retirement. This study reveals that these following variables: age, education level, healthcare eligibility, marital status, health condition, total assets, gender, residential type, percent of elderly in the household, and number of children have strong influences on an elderly's desire to continue work. By knowing which factors contribute to the elderly wish to continue work in the market, this research allows for future prediction of the labor market that can accommodate elderly in Thailand. Our final models of random forest and lasso logistic regression provide prediction accuracy of 68.19 and 69.58 percent on the elderly's desire to work, respectively. This study has a significant impact as policymakers can utilize our models in predicting elderly's desire to work after retirement age and design a labor market that can accommodate elderly in Thailand in the future.
\end{abstract}

\section{Introduction}
\label{sec:intro}

\subsection{Elderly} \label{sub:elder}
Countries around the world consider an individual to be an elderly at a different age. Elderly are divided into different stages, for example, 65-74, 75-84, and those above 85. According to the Thailand Act on the Elderly (2003), it determines that an elderly is an individual who is 60 or above (\citealt{ministry_of_social_development_and_human_security_act_2003}). 

Currently, the amount of elderly population is rising around the globe. By 2050, people aged 60 and above will reach 2.4 billion (\citealt{oecd_ageing_2015}). Declining fertility and increasing life expectancy are the main causes of aging society (\citealt{institute_for_population_and_social_research_situation_2019}; \citealt{phijaisanit_how_2016}). Thai society has entered into an aging society since the year 2000. By 2010, the population age 60 and above in Thailand amounts to 8.4 million which is 13.2 percent of the total population (\citealt{wattanasaovaluk_economic_2021}; \citealt{wongboonsin_aging_2017}). It is estimated that in 2021 Thailand will enter into a complete aged society, which means that Thai population 60 years and above will account for twenty percent of the total population and by 2031 Thailand will become a super-aged society with population 60 and above amounting to twenty-eight percent of the total population (\citealt{National_Statistical_Office}; \citealt{srisuchart_promotion_2019}).

\subsection{Retirement} \label{sub:retirement}
	As the elderly population in Thailand is on the rise, it is important that we examine the retirement trends. Retirement is an important topic to examine as the timing of retirement has a crucial impact on a country. It has an effect on the labor market, the social security payouts, the caregiving industry as well as the overall productivity of a country's economy (\citealt{arkornsakul_labor_2020}; \citealt{bloom_macroeconomic_2015}). 
	
Retirement for the elderly varies from country to country and also varies for men and women (\citealt{oecd_pensions_2013}). Moreover, it changes over time. In the United States, for instance, in 1950, men on average retire at 70, in 1970 it dropped to 65 and by 1985 it dropped to 62 (\citealt{quinn_changing_2002}). Across 34 countries in the Organization for Economic Co-operation and Development (OECD), men's average retirement age in 1949 was 64.3 years and in 1999 it decreased to 62.4 years, and in 2012 it increased to 64.2 years. Women's average retirement age in 1949 was 62.9 years and in 1999 it decreased to 61.1 years and in 2012 it increased to 63.1 years. (\citealt{oecd_pensions_2013}). In Thailand, the retirement age for the public sector is sixty years of age, however, for the private sector, the retirement age varies depending on the occupation and the business industry (\citealt{Office_of_the_Council_of_State}; \citealt{Srawooth_Paitoonpong_2018}; \citealt{Labor_Protection_Act_2017}).

Retirement is not a one-time discrete event; rather it is a process in which individuals transition from full employment to full retirement (\citealt{beehr_process_1986}; \citealt{denton_what_2009}; \citealt{wang_employee_2010}). Some workers engage in bridge employment, that is, individuals continue working for pay after they retire from a career job. Bridge employment can take multiple forms; for example, reduce work hours on a current job or move to a less demanding job (\citealt{beehr_working_2015}). Bridge employment is becoming a common phenomenon. Seventy percent of current workers plan to work for pay after retirement (\citealt{quinn_work_2010}). Other workers engage in unretirement which is the transition from full retirement back to part-time or full-time employment (\citealt{maestas_back_2010}). 

Elderly workers retire due to several reasons such as declines in health status, negative job conditions, family caregiving responsibilities, and desire for leisure pursuits (\citealt{fisher_retirement_2016}; \citealt{hansson_successful_1997}; \citealt{phijaisanit_how_2016}). One of the most cited reasons for early retirement is poor health (\citealt{mcgarry_health_2004}; \citealt{park_health_2010}; \citealt{van_rijn_influence_2014}). Health conditions that lead to early retirement include cardiovascular conditions (e.g. heart problems and stroke), musculoskeletal conditions (e.g. back pains), and mental illness (e.g. anxiety and depression) (\citealt{fisher_retirement_2016}).

\subsection{Do elderly want to work?} \label{sub:want}
Many countries around the world are now embracing the concept of “Active Aging.” Active Aging is a policy that support people as they age to remain in charge of their own lives and to encourage their continuing contribution to the economy and society. This concept underscores that the elderly should remain active participators in social, economic, cultural, and civic activities (\citealt{world_health_organization_active_2002}). One important implication of such a concept is the elderly desire to remain active in the labor market. Elderly should be able to participate in the job markets according to their individual needs and abilities (\citealt{world_health_organization_active_2002}).
Existing literature (e.g. \citealt{kubickova_active_2018}) indicates that the elderly decision to stay on in the labor market is due to various factors such as income, health condition, social contacts, government policies, and living conditions. 

Income and wealth are factors that determine one's decision to continue to work or retire. Some studies (e.g. \citealt{schils_early_2008}) find that workers with more accumulated wealth were more likely to retire as they feel financially secure and can choose between continue working or increase leisure time. Yet, other studies (e.g. \citealt{parker_retirement_2007}) find that individuals with higher incomes, due to high opportunity costs, may prefer to continue earning thus retiring later. \citet{gustman_employer_1994} find that workers delay retirement until the date at which they become eligible for health benefits after retirement and are more likely to retire once eligible. In the United States, for instance, eligibility for Medicare at age 65 is an important factor in retirement decisions. Workers who have employer-provided health insurance (EHI) but not retiree health insurance (RHI) were especially likely to retire when they turn 65 (\citealt{coe_how_2013}).

In Thailand, retirement pension and healthcare eligibility are two important factors that affect elderly's decision to remain in the job market. Thai elderly can expect to receive different types of monetary and health welfare depending on their life long career. Elderly can be categorized into three groups based on their past career 1) elderly who worked in a public sector 2) elderly who worked in a private sector and 3) elderly who worked as a freelance.

Elderly who worked in the public sector will receive a government pension and free access under certain conditions to public hospital until they die (\citealt{the_comptroller_generals_department_civil_2021}). Elderly who worked in the private sector have employer-sponsored pension plans and healthcare through Thailand Social Security Scheme (SSS). Yet, the healthcare coverage ends upon retirement. The most vulnerable group is the elderly who worked as freelance. Most freelance do not have pension plan and healthcare plan, rather they must rely on their own savings.\footnote{There are exceptions as some elderly do plan their retirement and set up private pension fund and buy their own private healthcare plan.} Further, many Thai elderly work in “informal” sector, agriculture, for instance, and there is no specific retirement age. Many elderly continue to work until they are not able to. 

While it is true that elderly in Thailand, as all Thai citizens with Thai national identification numbers, are entitled to a free healthcare program- a universal healthcare coverage often known as “Bat Thong,”\footnote{Some exclusion applies, for example, those who are eligible for other forms of healthcare services, such as government officials.} 
yet, this free healthcare program is widely known for its delay and inferiority in medical treatment. This could be an explanation as to why Thai elderly choose to remain in the labor market to continue receiving private healthcare provided by the employers.

As seen in Table \ref{tab:eldnum}, many Thai elderly continue to work. In 2012, 3,403,873 elderly worked and by 2020 4,704,477 elderly worked. The percentage of elderly being in the labor market is between 39.46 to 42.14 between 2012 and 2020, reflecting that more than a third of Thai elderly still hold a job in their retirement age (\citealt{national_statistical_office_demographic_2021}).

\begin{table}[H]
\begin{center}
 \includegraphics[width=5.5in]{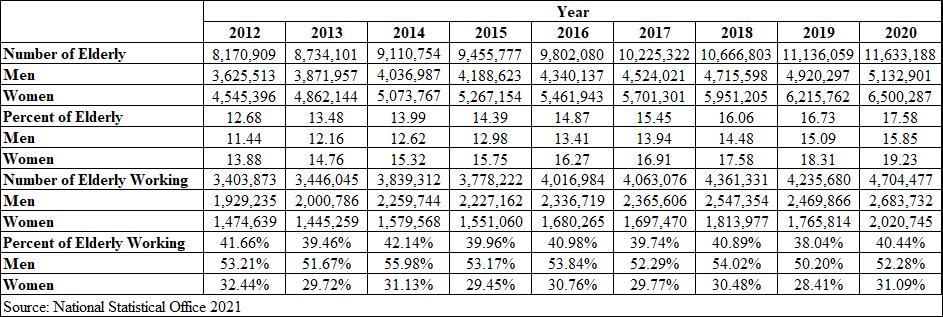}
\end{center}
\caption{Number and percentage of elderly in the labor market in Thailand}
\label{tab:eldnum}      
\end{table}

This article examines the factors that are associated with the elderly's decision to remain in the labor market after retirement age. Doing so, we are able to accurately forecast using statistical models which elderly want to continue working. This will allow us to design a labor market that can respond to the needs of the elderly in the near future.

\section{Data}
\label{sec:data}

\subsection{Sampling} \label{sub:sampling}

We are using data from the National Statistical Office, Ministry of Digital Economy and Society. The 2017 Survey of the Older Persons in Thailand is the sixth survey that is collected by National Statistical Office of Thailand regarding elderly in Thailand. The data was collected countrywide from June to August 2017. This sample survey uses a stratified two-stage sampling. There are 77 stratums based on the province, then each stratum is categorized as two sub-stratum-municipal areas and non-municipal areas. This leads to 5,970 enumeration areas which then results in a total of 83,880 households with a total of 217,818 respondents. We then intentionally pick observations of those who are above 60\footnote{{We exclude 60-year-old people as retirement in the public sector in Thailand considers the year of birth for retirement. Therefore, some 60-year-old may still be working.}}, this leads to a total of 38,551 observations.  After taking into account missing values, we are left with 31,190 observations. 

The 2017 Survey of the Older Persons in Thailand collects data on demographic, economic, social and health characteristics, and household types of elderly. The data has 186 variables that are categorized into ten groups: identification, general information, children status, job status, income and assets, living condition, support from children, health condition, participation in elder activities, caretaker, and residential condition. We predict the elderly's desire to work (yes/no) based on these fourteen variables: age, gender, education level, marital status, number of children, number of family members in the household (excluding him/herself), number of grandchildren in the household, percentage of elderly in the household, children contact frequency, total assets\footnote{Total assets variable was selected over income variable as most elderly who do not want to work have zero income. Therefore, if the income variable was selected the models will be very accurate yet other factors will not be detected.}, health condition\footnote{Health condition ranges from 1 to 5. One indicates that the elderly is in a very bad health condition, and five indicates that the elderly is in a very good health condition. The health condition is based on the elderly's own evaluation.}, happiness level\footnote{Happiness level ranges from 0 to 10. Zero indicates that the elderly is depressed and ten indicates that the elderly is exuberant. The happiness level is based on the elderly's own evaluation.}, healthcare eligibility, and residential type.

From the 31,190 observations, the number of elderly who desire to work is 8,885 or 28.49 percent. Training the model on the 31,190 observations, the classification model predicts all observations as “no desire to work.” This model has a 28.49 percent error; however, we did not further investigate it as it does not provide useful information and it is not useful in making prediction. To understand how the fourteen variables influence the elderly's decision to work, we intentionally undersample the elderly who do not wish to work by uniformly choosing 8,885 in this class from the 22,305 observations. The final sample has 17,700 observations consisting of 8,885 who desire to work and 8,885 who do not desire to work. We use the undersampling technique rather than the oversampling technique such as Synthetic Minority Over-sampling Technique (SMOTE) as the sample is large and most of the variables are categorical. Since our undersampled data has an equal fifty percent of elderly who wish and do not wish to work, our model will not be biased towards each outcome. This then leads to effective models in term of prediction and important variable selection.\footnote{The undersampled data is used in lasso logistic regression and random forest analysis. The pre-undersampled data with 31,190 observations is used in a cross tabulation analysis.}

\subsection{Response and predictors}

The response variable is the desire to work of each elderly which can be categorized as ``does not desire to work'' and ``desire to work.'' The variable is a binary response of 0 and 1, respectively. The original data before undersampling consists of 28.49 percent of elderly who desire to work and 71.51 percent who do not desire to work.
The variables predicting the desire to work for this study include:
\begin{enumerate}
	\item Age ($x_1$)
	\item Gender ($x_2$)
	\item Education level ($x_3$)
	\item Marital status ($x_4$)
	\item Number of children ($x_5$)
	\item Number of family members in the household ($x_6$)
	\item Number of grandchildren in the household ($x_7$)
	\item Percent of elderly in the household  ($x_{8}$)
	\item Children contact frequency ($x_{9}$) 
	\item Total assets ($x_{10}$) 
	\item Health condition ($x_{11}$)
	\item Happiness level ($x_{12}$)
	\item Healthcare eligibility ($x_{13}$)
	\item Residential type ($x_{14}$)
\end{enumerate}

Previous studies, for example, \citet{arkornsakul_labor_2020}, \citet{bai_financial_2020}, \citet{coe_how_2013}, \citet{kim_factors_2016}, \cite{matthews_family_2013}, \citet{turner_factors_1994} and \citet{wattanasaovaluk_economic_2021}, utilize only some of these mentioned variables. We strengthen our study by incorporating all of these variables and using a different statistical approach to see how each variable influences the elderly's desire to work.

\begin{table}[H]
\begin{center}
 \includegraphics[width=6in]{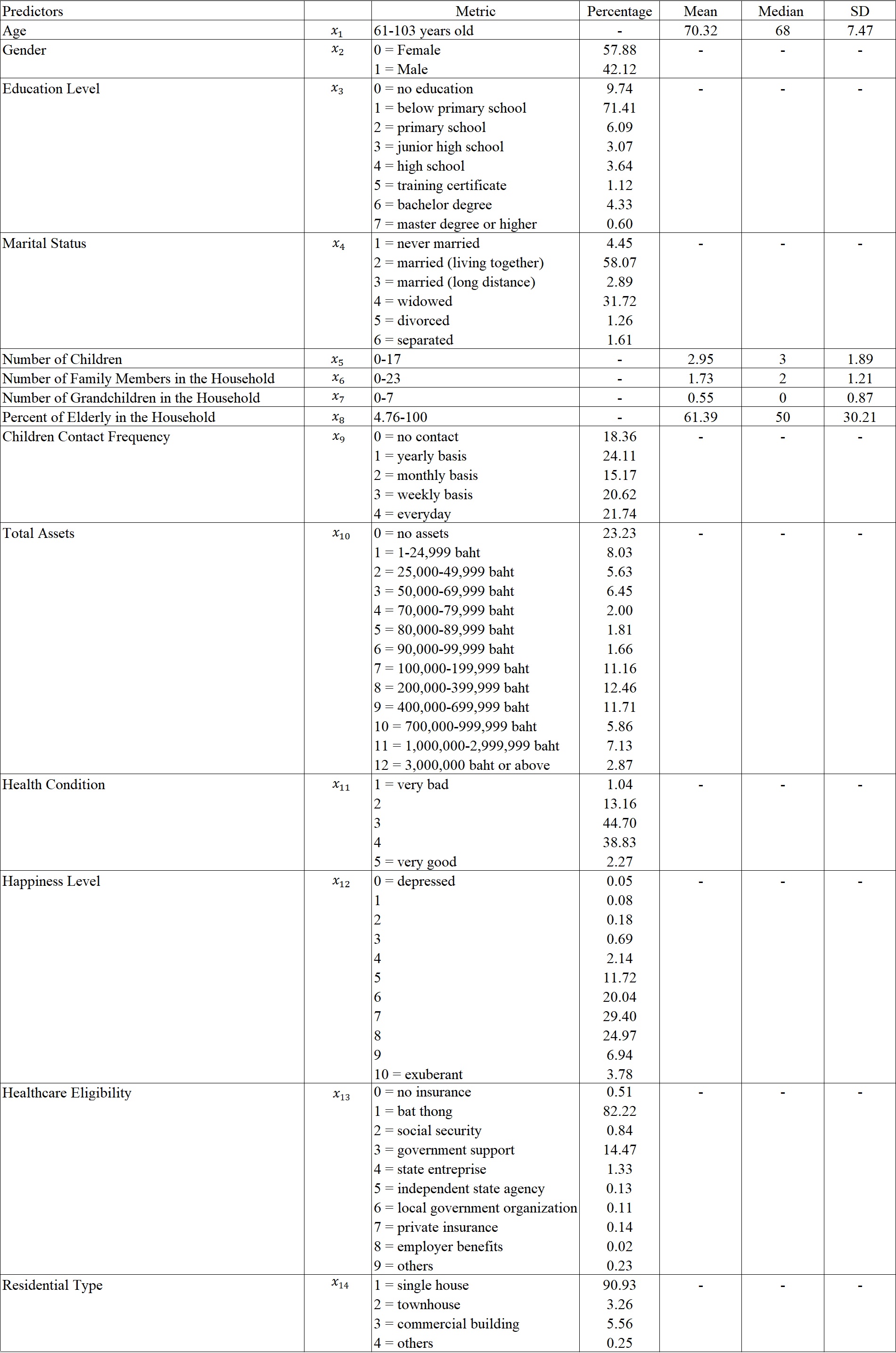}
\end{center}
\caption{Predictor Descriptions, Metrics, Means, Medians, and Standard Deviations }
\label{tab:descriptive}      
\end{table}

In Table \ref{tab:descriptive}, the variables are listed. The variables $x_1,x_5,x_6,x_7$ and $x_8$ are quantitative. The variables $x_3, x_9,x_{10},x_{11}$ and $x_{12}$ are collected as categorical but are orderable as detailed. The nominal variables include $x_2, x_4, x_{13}$ and $x_{14}$. The descriptive statistics of all the predictors are shown in Table \ref{tab:descriptive}. For quantitative variables, the mean, median and SD are shown and for categorical variables, the categorical percentages are shown. 

Note that this dataset contains the current job and wage of the elderly who is still working but not their prior jobs. Therefore, we do not include these variables as we consider all elderly in the dataset. Based on the dataset, the elderly who are still working mostly still want to work.


Next we report the generalized variance inflation factors (GVIF) (\citealt{fox1992generalized}) to show that these predictors have no colinearity issue. Recall that GVIF is the generalized version of the variance inflation factor (VIF) that can be used for categorical variable. We consider $\left(\texttt{GVIF}^{1/(2\cdot \texttt{DF})}\right)^2$ which can use the same rule of thumb as VIF. The values close to 1 indicates no colinearity issue and the values greater than 5 indicate too much colinearity. The GVIF and $\left(\texttt{GVIF}^{1/(2\cdot \texttt{DF})}\right)^2$ of the 14 predictors in Table \ref{tab:vif} below show no sign of colinearity.

\begin{table}[h] 
\renewcommand{\arraystretch}{0.7}
\caption{The GVIF table.}
\centering
\scriptsize	
\resizebox{\columnwidth}{!}{%
 \begin{tabular}{|c| c c c c c c c c c c c c c c |} 
 \hline
 m   & $x_{1}$ & $x_{2}$ & $x_{3}$ & $x_{4}$ & $x_{5}$ & $x_{6}$ & $x_{7}$ & $x_{8}$ & $x_{9}$ & $x_{10}$ & $x_{11}$ & $x_{12}$ & $x_{13}$ & $x_{14}$  \\ [0.5ex] 
 \hline \\ [-0.3ex]
 GVIF	& 1.18 &  1.21 & 1.45 & 2.40 & 1.59 &  3.41 & 1.84 & 2.82 & 1.73 &  1.07 & 1.28  &   1.29   &    1.36   &    1.08 \\ [0.5ex] 
 DF	& 1 &  1  & 1 & 5 & 1 & 1 &  1 & 1 &  4  & 1  & 1 &  1 & 10 & 3 \\ [0.5ex] 
 $\left(\texttt{GVIF}^{1/(2\cdot \texttt{DF})}\right)^2$	& 1.18 &  1.21 & 1.45 & 1.19 & 1.59 &  3.41 & 1.84 & 2.82 & 1.15 &  1.07 & 1.28  &   1.29   &    1.03   &    1.03  \\ [1ex] 
 \hline
\end{tabular}%
}
\label{tab:vif}
\end{table}


While many researchers focus on work participation of the elderly and the factors that influence the elderly's choice, yet most use either linear regression or ordinary logistic regression (for example \citealt{haider_elderly_2001}; \citealt{jean-olivier_distance_2010}; \citealt{kubickova_active_2018}), rather than shrinkage logistic regression (lasso logistic regression) and decision tree (random forest).

Lasso logistic regression is widely used in science, especially in the medical field (for example, see \citealt{dai_use_2016} and \citealt{lee_application_2014}) but it is not the case in social science. Lasso logistic regression has many advantages over other logistic regressions as it reduces the variation of the model and allows for variable selection. Several social science studies have begun to incorporate this method (for example see \citealt{elyasiani_determinants_2019}).

A study by \citet{molee_study_2019} had attempted to use decision tree (random forest) in studying elderly, however, their study was focused on the risk of catching a disease among the elderly. The decision tree's main advantage is that it is easily visualizable and random forest provides the ranking of the variable importance.

\section{Methods} \label{sec:method}

\subsection{Cross tabulation}

Prior to analyzing the data using lasso logistic regression and random forest, cross tabulation is used to visualize the relationship between each variable and the desire to work based on pre-undersampled data of 31,190 observations.\footnote{The results from the cross tabulation is merely for preliminary analysis. No conclusive results is established in this paper based on the cross tabulation as there could be a correlation between the predictors. For example, in a cross tab, a widower may show as not wanting to work, yet the actual reason could be that most widowers have the tendency to be older than other elderly.} Note that we do not report the chi-square test of independence as our sample is very large and thus any small difference will appear statistically significant.

\subsection{Random forest}


The idea of random forest was first introduced by \citet{ho1995random} and an extension of it which was termed random forest was registered as a trademark in 1998 by Breiman and Cutler (see \citealt{breiman2001random} or \citealt{james_introduction_2013} for reviews of the methods). We apply random forest with the undersampled data of 17,700 observations to obtain models to predict the elderly desire to work. Random forest is one of the tree-based methods that separates the predictors space into sub-regions where each region predicts the same outcome. Random forest is used to overcome the weakness of the basic decision tree that has a very high variance by generating several basic decision trees and makes a prediction towards the majority of the outcomes in each sub-regions from the trees. Random forest also allows us to select the number of variables to include in each tree and by selecting a smaller random tree in each step it will help detect the importance of each variable without biasing towards only the most important one. The most popular choice of the number of variables selected for each tree is $\sqrt{p}$ where $p$ is the total number of variables. In this research, we generate 100 trees and select four variables randomly for each tree. For an observation, each tree predicts 0 or 1 and we take the majority of either 0 or 1 as the final outcome. 

In this work, trees are grown based on Gini index which is defined as
\beas
G=\hat{p}_{m0}(1-\hat{p}_{m0})+\hat{p}_{m1}(1-\hat{p}_{m1})=2\hat{p}_{m1}(1-\hat{p}_{m1})
\enas
where $\hat{p}_{m0}$ and $\hat{p}_{m1}$ denote the proportions of training observations in the $m^{th}$ region that are from class 0 and 1, respectively. It is clear that the values of $\hat{p}_{m1}$ close to either 0 or 1 will result in a small Gini index. The random forest algorithm also provides the importance ranking of predictors. The well-known importance ranking is based on either mean decrease Gini (MDG) or mean decrease accuracy (MDA). MDG of each predictor is defined as the difference between the means of Gini index from all regions separated by the random forest model with and without that predictor. MDA is defined similarly with Gini index replaced by predicting accuracy. If the model without a particular predictor results in a much larger Gini index or much lower accuracy, then MDG and MDA of this predictor are large which imply that the predictor is important. Nevertheless, MDG and MDA do not necessarily lead to the same outcome. A variable may has a good MDA but not a good MDG, vice versa.

\subsection{Lasso logistic regression}


The idea of lasso was introduced by \citet{santosa1986linear} and later \citet{tibshirani1996regression} coined the term ``lasso.''  Binary logistic regression model (see \citealt{hastie_elements_2009} for review of the methods) is of the form
\beas
\log\left(\frac{P(Y=1)}{P(Y=0)}\right) = \beta_0+\beta_1X_1+\beta_2X_2+\cdots+\beta_pX_p
\enas
where $P(Y=0)$ and $P(Y=1)$ denote the probability that the outcome is 0 and 1, respectively, and $X_1,X_2,\ldots,X_p$ denote $p$ predictors. The method solves for $\beta_j$ that maximize the binomial likelihood function
\beas
L(\beta_j) = \prod_{j:y_j=1} P(Y=1|\vec{x}_j) \prod_{j:y_j=0} P(Y=0|\vec{x}_j)
\enas
where 
\beas
P(Y=1|\vec{x})=\frac{e^{\beta_0+\beta_1 x_1+\beta_2 x_2 + \cdots +\beta_p x_p}}{1+e^{\beta_0+\beta_1 x_1+\beta_2 x_2+\cdots+\beta_p x_p}}
\enas
and
\beas
P(Y=0|\vec{x})=\frac{1}{1+e^{\beta_0+\beta_1 x_1+\beta_2 x_2+\cdots+\beta_p x_p}}.
\enas
Note that the model above is the same if we minimize $-L(\beta_j)$ instead. Lasso logistic regression is the model that solves $\beta_j$ that minimize
\bea \label{lassoMLE}
-L(\beta_j) + \lambda \sum_{j=1}^p|\beta_j|
\ena
with $\lambda\ge 0 $ as a tuning parameter. The second term is called the penalty term and it will force $\beta_j$ to shrink towards zero as $\lambda$ is larger. Therefore, if we can seek an appropriate $\lambda$, it will reduce the variance of the logistic regression model and will select only the most important factors in the final model. We use this model as a classification tool by assigning an individual with $P(Y=1|\vec{x})>0$.5 or 50\% to those who correspond to ``have a desire to work'' throughout the paper.  We remark here that in some applications it is possible that 50\% is not the best threshold. Logistic regression uses the undersampled data of 17,700 observations to obtain models to predict the elderly desire to work.  

\subsection{$k$-fold cross validation}

In this work, we apply $k$-fold cross validation to each model in order to compare its accuracy. This method first splits the original dataset into $k$ nonoverlap sets. Then each time for $k$ times it constructs the model based on $k-1$ sets and leaves one set for testing accuracy. The $k$-fold cross validation accuracy is the average accuracy from $k$ times. The popular choice of $k$ in machine learning is $10$. We thus use $10$ throughout this work. This method has a main advantage in that it provides an accurate accuracy as it reduces the variance of the estimate accuracy by repeating the process $10$ times and taking the average. 

In lasso logistic regression, we apply $10$-fold cross validation to compare the models with different tuning parameters $\lambda$ in order to select an appropriate $\lambda$. We also use $10$-fold cross validation technique to compare the final models from random forest and lasso logistic regression based on their accuracy.

\section{Results and discussions}
\label{sec:result}

\subsection{Cross tabulation}

\begin{figure}[H]
\begin{center}
 \includegraphics[width=3.7in]{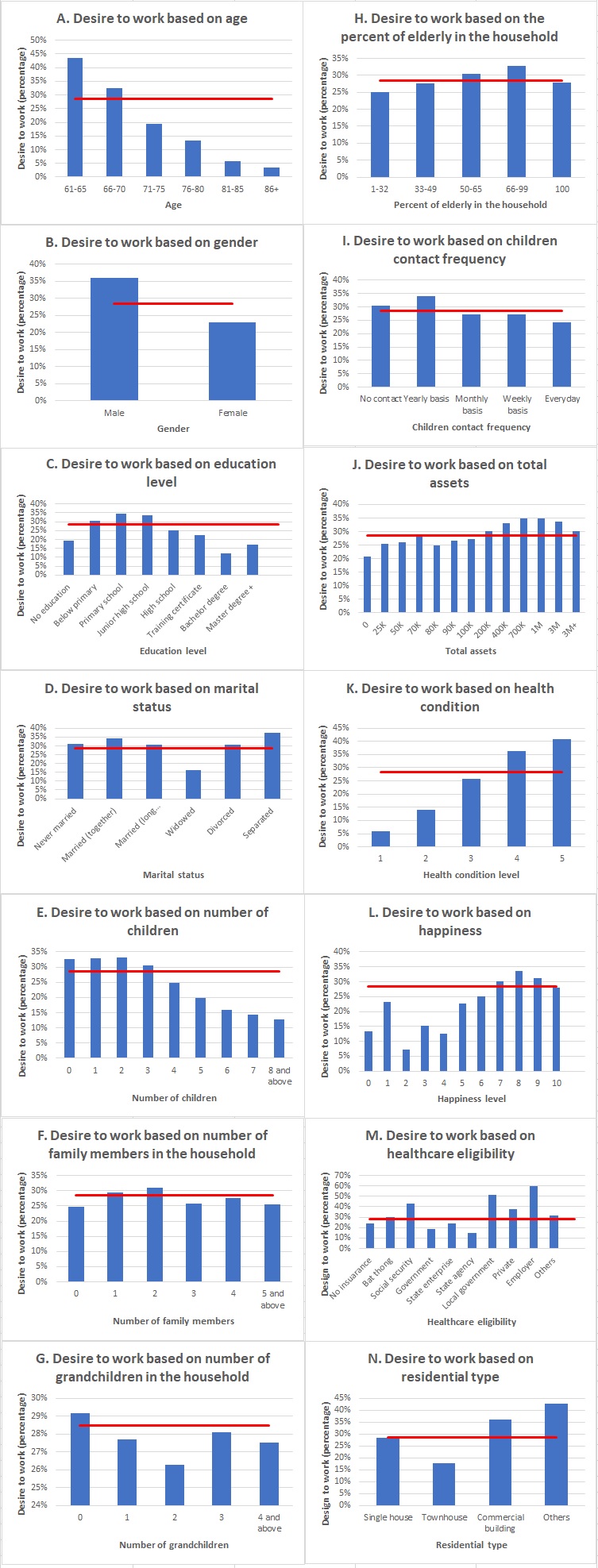}
\end{center}
\caption{Cross tabulation bar graphs. (The blue column represents the percentage of elderly who desire to work in each category. The red horizontal line represents the overall percentage of elderly who desire to work.)}
\label{fig:bar}      
\end{figure}

First, we perform cross tabulation analysis as a preliminary evaluation of the relationship between each of the fourteen predictors and the elderly's desire to work. The results are shown in Table \ref{tab:cttable}. We also plot bar graphs in Figure \ref{fig:bar} which illustrate the percentage of elderly who desire to work in each category of each predictor comparing to the average of all observations.

Figure \ref{fig:bar}(A) shows the obvious, that is, as one ages the desire to work decreases. This is probably due to the physical stamina and the health condition of the elderly. This finding is in line with \citeauthor{wattanasaovaluk_economic_2021}'s study (\citeyear{wattanasaovaluk_economic_2021}) in that elderly exit the labor market because they are too old to work.  

As Table \ref{tab:eldnum} illustrates, in the past few years, elderly’s intentions to work have been increasing for both genders (\citealt{national_statistical_office_demographic_2021}.) Our study, Figure \ref{fig:bar}(B) shows that elderly men are more likely than elderly women to continue work after retirement. This is consistent with \citeauthor {arkornsakul_labor_2020}’s study (\citeyear{arkornsakul_labor_2020}) which claims that men are considered to be the head of the household thus the need to earn money to take care of the family and men are physically stronger than women which encourage men to stay in the labor market longer than women.

For the education level variable (Figure \ref{fig:bar}(C)), it shows that elderly with no education and those with more than high school education have less tendency to want to work than the average. Only elderly with below primary school, primary school, and junior high as their highest educational attainment have the desire to work more than the average pool. Our findings is consistent with that of \citeauthor{kim_factors_2016}'s research (\citeyear{kim_factors_2016}) and \citeauthor{arkornsakul_labor_2020}'s study (\citeyear{arkornsakul_labor_2020}) which show that elderly workers with a higher education level were less likely to maintain employment. This perhaps is because those with high educational attainment are more likely to have a well-paid job during their working years and thus have enough savings to retire. On the other hand, elderly with no education at all are mostly very old and thus have no desire to work.\footnote{In 1921, the Compulsory Education Act required that every child from seven to fourteen attend schools. Throughout 1921 to 1960, the number of years of compulsory education increased however it was not strictly enforced. From 1960 to 1978, four years of primary education was required and in 1978, six years of primary schooling was made compulsory (\citealt{curran_boys_2002}; \citealt{keyes_proposed_1991}).}

Figure \ref{fig:bar}(D) shows that married elderly and elderly who are separated are more likely to have the desire to be in the labor market. This is because elderly who are widowers tend to be much older than those of other marital status, hence the old age prevents one from working. As for other marital status categories, the total count is too small to have any statistical significance.

Number of children plays an important role in whether or not an elderly wants to continue working. Figure \ref{fig:bar}(E) shows that the more children an elderly has, the more likely he/she does not wish to work. This possibly reflects Thai culture that children often give their parents money as they age. Hence, elderly with lots of children may be able to cover the entire expenses without the need to work (\citealt{knodel_familial_1992}; \citealt{knodel_intergenerational_2008}; \citealt{phijaisanit_how_2016}).

\citeauthor{arkornsakul_labor_2020}'s study (\citeyear{arkornsakul_labor_2020}) finds that the elderly's intention to continue work depends on the number of household members. This is the case for agricultural careers as it requires a lot of workforces and physical effort. If there is a decrease of household members, the elderly will decide to continue working as to maintain their income level. While our study did not dissect studying elderly at a regional level as \citeauthor{arkornsakul_labor_2020}'s study (\citeyear{arkornsakul_labor_2020}), yet the big picture in Thailand as Figure \ref{fig:bar}(F) illustrates, elderly who has one or two family members in the household excluding the elderly themselves have the tendency to want to work more than the average pool. This possibly reflects that elderly may not have enough earnings to upkeep the household and thus need to continue working (\citealt{phijaisanit_how_2016}). Alternatively, it could be that these elderly choose to keep their employment as it maintains their well-being and ensures that they can support themselves in the future (\citealt{srisuchart_promotion_2019}). Elderly with many family members have less desire to work than the average pool, this perhaps is because each family members help pitch in to make ends meet. Interestingly, elderly who live by themselves are less likely than other elderly to want to continue work. This could be that they no longer need to financially support anyone but themselves thus they do not see the point of holding a job. This contradicts with \citeauthor{santiphop_analysis_2016}'s study (\citeyear{santiphop_analysis_2016}) which finds that elderly who lived alone have to remain employed as it may be equivalent to survival.

Figure \ref{fig:bar}(G) shows that elderly with no grandchildren have the highest desire to continue working. This reflects that these elderly do not need to help pitch in to childcare and thus can decide to work. On the other hand, elderly with any amount of grandchildren have less than the average desire to work as they may be asked to help out with childcare. Globally, and especially so in Thailand, it is quite common for children to ask their parents to help take care of their kids while they are at work. Some grandparents take care of their grandchildren full time (\citealt{hank_grandparents_2009}; \citealt{hayslip_jr_grandparents_2005}; \citealt{kamnuansilpa_grandparents_2005}; \citealt{komonpaisarn2019providing}; \citealt{mehta_introduction_2012}; \citealt{nanthamongkolchai_physical_2012}).

Figure \ref{fig:bar}(H) shows that in a household with most members being seniors, but not all seniors, more than an average tend to want to work. Perhaps, living with other elderly they still need to work as they do not have enough money to pay for their expenses and maybe they need to support other dependents within the households as well. However elderly in households with fewer elderly or households with one hundred percent elderly may have other family members support their cost of living or may not need to support dependents, respectively. This is consistent with \citeauthor{arkornsakul_labor_2020}'s study (\citeyear{arkornsakul_labor_2020}) and \citet{bank_of_thailand_aging_2018} which indicated that elderlies leave the labor market as they have to take care of the family members who are children, seniors, and/or sick.

Activity theory considers the more activities and the more socialization that an elderly gets, the happier the elderly will be (\citealt{havinghurst_patterns_1968}; \citealt{little_aging_2016}). Children's contact frequency thus plays an important role in determining whether or not the elderly wish to work. Figure \ref{fig:bar}(I) demonstrates that elderly who have no contact with their children or have few contacts with their children per year still have the desire to work more than the average. Work for these elderly may be a way for the elderly to keep themselves busy and/or is their source of happiness. Elderly that have frequent contact with their children on the other hand are more likely to want to retire because they are able to seek a new source of happiness that is doing activities and spending time with their loved ones.

Continuity theory explains that elderly do not change their ways of living in a sudden manner, rather they progressively make a rational choice about their future based on their social roles (\citealt{atchley_continuity_1989}; \citealt{von_bonsdorff_continuity_2013}). Figure \ref{fig:bar}(J) shows that the percentage of elderly with no asset or little to moderate asset (0 to 100,000 baht) has the desire to work below average. This conceivably is because there is no motivation in working as the compensations from work is rather low. The percent of elderly with lots of assets (above 200,000 baht), on the other hand, has the desire to work above average. This could be due to the high monetary returns. The figure further shows that there is an obvious peak at 700,000-1,000,000 baht. Due to data restriction which did not differentiate elderly with more than three million baht assets, we cannot directly infer the wealthy trends towards work desires. Nevertheless, due to the decreasing trend thereafter we believe that people with very high assets will have less desire to work.

\citet{santiphop_analysis_2016} explain health condition as one of the main factor that affects elderly's employment status. Our study (Figure \ref{fig:bar}(K)) shows that elderly with better health are more likely to remain in the labor market. This is consistent with \citeauthor{haseen_self-assessed_2010}'s study (\citeyear{haseen_self-assessed_2010}) which finds that elderly who did not work were more likely to report poor health than those who worked.

\begin{table}[H]
\begin{center}
 \includegraphics[width=4.3in]{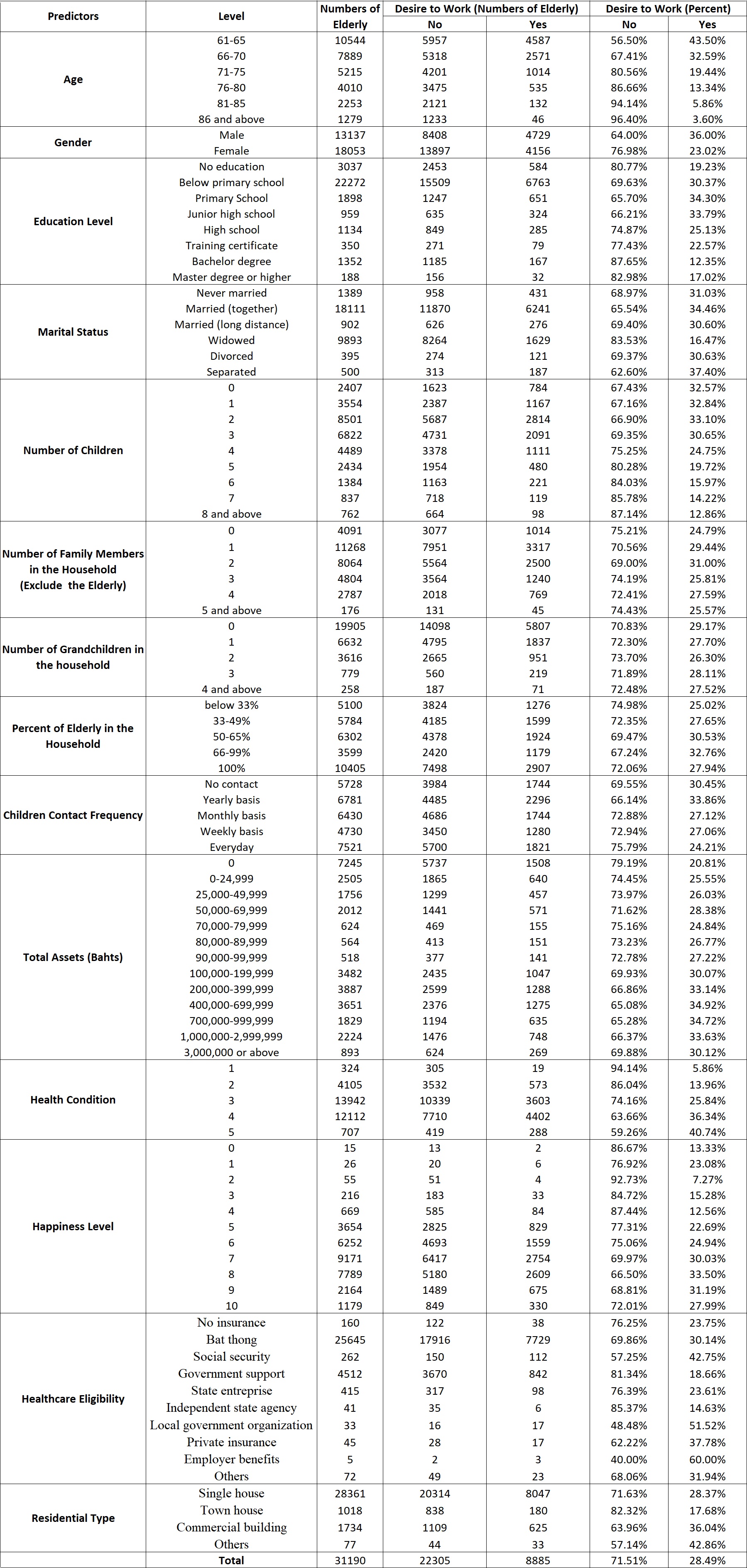}
\end{center}
\caption{Cross tabulation table}
\label{tab:cttable}      
\end{table}

Figure \ref{fig:bar}(L) shows that elderly who evaluate themselves as depressed do not want to work, whereas those who are happy tend to work more than the average. The figure shows that the elderly desire to work peaks at happiness level 8, this possibly reflects that those who still want to work, decide to work and thus are happy. Interestingly, those who are extremely happy (above 8) have less tendency to want to work. This may be that their happiness is not contingent only upon work but their means to happiness can come from other sources (\citealt{chyi_determinants_2012}; \citealt{gray_inner_2008}; \citealt{kehn_predictors_1995}; \citealt{nanthamongkolchai_physical_2012}).

As mentioned earlier, eligibility for healthcare plays a great role in elderly’s decision to remain in the job market. Thai elderly receive different types of healthcare upon retirement depending on their life long career. In Figure \ref{fig:bar}(M), we can see two separate trends. Elderly who worked for the government are more likely to have less desire to continue work after retirement and this is precisely because working for the government elderly will receive healthcare coverage even upon retirement (\citealt{the_comptroller_generals_department_civil_2021}). On the other hand, elderly who worked in a private sector their healthcare coverage stop once they retire. Therefore, to have access to healthcare many elderly who worked in a private sector intentionally choose to continue work.

Figure \ref{fig:bar}(N) shows that elderly who live in single houses and town houses have less desire than the average to continue work. On the other hand, elderly who live in commercial buildings have the desire to continue work. These elderly may explicitly choose to live in such residential type as they can continue their work, selling merchandises, for instance.

\subsection{Random forest}

\begin{figure}[H]
\begin{center}
 \includegraphics[width=5.8in]{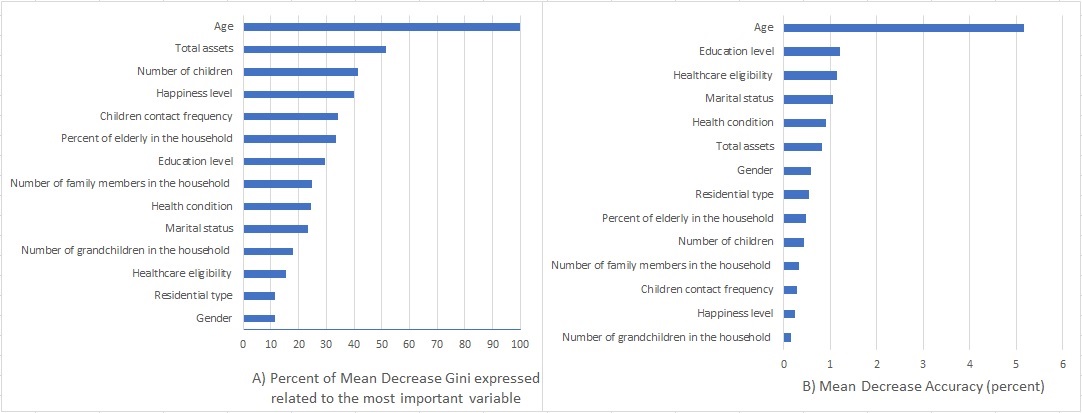}
\end{center}
\caption{Variables ranking in predicting the elderly's decision to work based on mean Gini index decrease (A) and mean decrease accuracy (B)}
\label{fig:imp}      
\end{figure}

We first apply the random forest technique that grows 500 trees with 4 as the number of random predictors for each tree. The random forest model  results in a 10-fold cross validation accuracy of 68.19 percent.  Figure \ref{fig:imp} shows the variables and their importance in predicting the elderly's decision to work based on MDG and MDA. \citet{strobl2007bias} finds that MDG is biased towards continuous variables and categorical variables with more numbers of categories. Since our predictors include both quantitative and categorical with different number of categories, MDA is more appropriate. As illustrated in Table \ref{tab:cttable}, for instance, men clearly have more desire to work than women but MDG suggests that this variable is the least important. Although these following predictors, age, education level, healthcare eligibility and marital status have MDA above 1 percent which have noticeably logical trends in the cross tabulation results in Figure \ref{fig:bar}, yet no conclusion can be made that the two results are consistent as random forest does not provide any directional impact information.

\subsection{Lasso logistic regression}

While random forest model provides the importance ranking of each variable, yet it has two disadvantages. First, random forest model does not show the direction of its impact. Second, it does not show which category of the categorical variables has the most impact. To overcome this weakness, we perform lasso logistic regression and compare the accuracy of the two models.

From Figure \ref{fig:imp}, children contact frequency, happiness level and number of grandchildren in the household are less impactful predictors (below 0.3 percent based on MDA) of the elderly's desire to work regarding the random forest model. We first remove them before performing lasso logistic regression. Note that we actually did some experiment and found that including these three variables in the lasso logistic regression model will lower the 10-fold cross validation accuracy by about 2 percent.

\begin{figure}[h]
\begin{center}
 \includegraphics[width=4in]{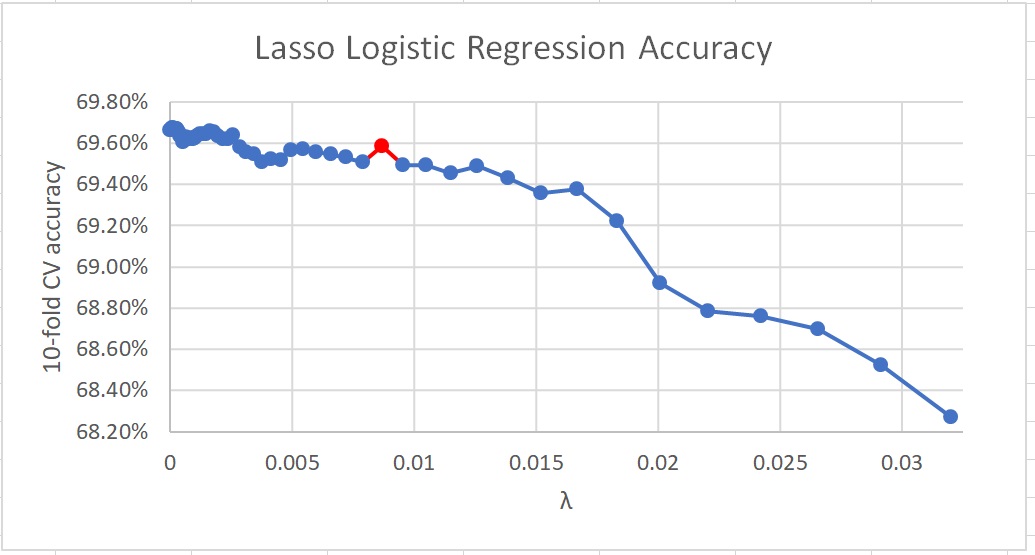}
\end{center}
\caption{Lasso logistic regression accuracy}
\label{fig:lambda}      
\end{figure}

Next, we apply lasso logistic regression after we scaled\footnote{Scaled means subtracting the mean and dividing by the standard deviation.} all numeric variables. We create a sequence of tuning parameter $\lambda$, as stated in \eqref{lassoMLE}, of the form $\lambda=10^k$ where $k$ is from $-4$ to $0$ and seek for an appropriate $\lambda$. Assigning an individual with $P(Y=1|x)>0.5$ to those who ``have a desire to work'', it turns out that the model with $\lambda=0.0001$ is the best model based on the 10-fold cross validation an accuracy of 69.67 percent. The coefficients of all the predictors from this model are nonzero which departs from our intention in using lasso logistic regression to also select important predictors. We thus intentionally move to the local peak as seen in Figure \ref{fig:lambda} based on accuracy at $\lambda = 0.0087$ which yields the 10-fold cross validation accuracy of 69.58 percent. The logit equation is
\beas
\log\left(\frac{P(Y=1)}{P(Y=0)}\right) &=& 5.819-0.697x_1+0.477x_{2,1}-0.252x_3+0.031x_{4,2} \\
          && \hspace{1pt} -0.386x_{4,4}-0.054x_5+0.186x_{10}+0.243x_{11} \\
					&& \hspace{1pt}-0.436x_{13,3}+0.034x_{13,7}-0.345x_{14,2}+0.295x_{14,3} .
\enas
where $x_{a,b}$ equals to 1 if the predictor $x_a$ is in category $b$ and 0 otherwise. 
The coefficients ranked by the sizes of their coefficients are summarized in Table \ref{tab:coef} below.

\begin{table}[h] 
\renewcommand{\arraystretch}{0.7}
\caption{Lasso logistic regression coefficients}
\centering
\scriptsize	
 \begin{tabular}{|c|c|} 
 \hline 
 \textbf{Intercept}   & \textbf{Coefficient} \\ [1ex] 
\hline 
  Intercept   & 5.819 \\ [1ex]
 \hline 
 \textbf{Positive predictor}   & \textbf{Coefficient}  \\ [1ex] 
\hline 
  Gender: Male  & 0.477  \\ [1ex] 
	
	Residential type : commercial building   & 0.295  \\ [1ex] 
	
	Health condition  & 0.243  \\ [1ex] 
	
	  Total assets    & 0.186  \\ [1ex] 
	
	Healthcare eligibility : private insurance  & 0.034  \\ [1ex] 
 
	Marital status: married (living together)  & 0.031  \\ [1ex] 

 \hline 
\textbf{Negative predictor}   & \textbf{Coefficient}     \\ [1ex] 
 \hline 
   Age & -0.697 \\ [1ex] 
	
	   Healthcare eligibility : government support  & -0.436  \\ [1ex] 
		
	 Marital status: widowed  & -0.386  \\ [1ex] 

   Residential type : townhouse & -0.345 \\ [1ex] 

   Education level & -0.252 \\ [1ex]

   Number of children & -0.054 \\ [1ex]

 \hline
\end{tabular}
\label{tab:coef}
\end{table}

From Table \ref{tab:coef}, there are six predictors that have positive impacts on the desire to work which are gender (male), residential type (commercial building), health condition, total assets, healthcare eligibility (private insurance), and marital status (married(living together)). On the other hand, the following predictors: age, healthcare eligibility (government support), marital status (widowed), residential type (townhouse), education level, and number of children, have negative impacts on the desire to work. The results here are consistent with that of the cross tabulation on the directional impact towards the desire to work. Yet, as shown in Table \ref{tab:coef}, lasso logistic regression goes a step further as it allows us to understand not only in term of prediction but also the ranking of factors that have influence on the elderly's decision to work. The larger coefficient magnitude means the more influence it has on the elderly's decision to work.  

We remark here that marital status (never married), healthcare eligibility (no insurance) and residential type (single house) are baseline categories for the multiple categorical variables which affect the intercept of 5.819. We cannot conclude the actual impact of these baseline categories, nevertheless, we know from the cross tabulation result in Table \ref{tab:cttable} that they have positive, negative and neutral impact, respectively.

\section{Conclusion and implication}
\label{sec:discussion}

Thai society has entered into an aging society since the year 2000 and is predicted to become a super-aged society by 2031 (\citealt{National_Statistical_Office}; \citealt{srisuchart_promotion_2019}; \citealt{wattanasaovaluk_economic_2021}). This change in the country's population demographic has important implication to the country in various dimensions, for example, the caregiving industry, the labor market, the social security payouts and the country’s economy productivity (\citealt{arkornsakul_labor_2020}; \citealt{bloom_macroeconomic_2015}). While many elderly consider retirement, yet another group of elderly choose to continue working. Using the 2017 Survey of the Older Persons in Thailand, this quantitative study uses cross tabulation, random forest with variable importance measure and lasso logistic regression to highlight factors that are associated with the elderly's decision to remain in the labor market and to build predictable elderly labor market models to respond to the aging society.

While previous literature (see \citealt{arkornsakul_labor_2020}, \citealt{bai_financial_2020}, \citealt{coe_how_2013}, \citealt{kim_factors_2016}, \citealt{matthews_family_2013}, and \citealt{turner_factors_1994}, for instance,) have examine factors that influence elderly's decision to retire, yet our study advances the literature by incorporating more variables from different studies. Doing so makes our research more robust. Additionally, we are able to focus on which factors to take into consideration when we implement policy regarding elderly labor market.

Another key difference from previous research is the use of random forest and shrinkage logistic regression. Random forest provides the variable importance ranking. Using random forest, our study shows that age, education level, healthcare eligibility, and marital status have the most impact on elderly's decision to remain in the market, with MDA of above 1 percent. Health condition, total assets, gender, residential type, percent of elderly in the household and number of children have some impact in elderly's decision-making with MDA of above 0.4 percent. Yet, as random forest model does not show the direction of its impact and does not show which category of the categorical variables has the most impact, we overcome this weakness by using lasso logistic regression. 

While lasso logistic regression is widely used in science, it is not the case in social science study. In the past, research on elderly's work participation and the factors that influence the elderly’s choice use either linear regression or ordinary logistic regression (for example \citealt{haider_elderly_2001}; \citealt{jean-olivier_distance_2010}; \citealt{kubickova_active_2018}). Lasso logistic regression has many advantages over other logistic regressions as it reduces the variation of the model and allows for variable selection. Using lasso logistic regression, our study illustrates that these following variables: gender (male), residential type (commercial building), health condition, total assets, healthcare eligibility (private insurance) and marital status (married and living together), have a positive impact on the elderly's decision to work while these variables: age, healthcare eligibility (government support), marital status (widowed), residential type (townhouse), education level and number of children, have a negative impact. This is consistent with the cross tabulation as all the positive (negative) impact quantitative and ordinal variables tend to have upward (downward) trends in the cross tabulation results while the percentages of elderly that still want to work in these positive (negative) impact categories in the nominal variable are above (below) average.

We are able to achieve the ambitious goal in predicting the elderly's decision to work with 68.19 and 69.58 percent accuracy, respectively by using random forest and lasso logistic regression models. By understanding which factors contribute to the elderly wish to continue work, policymakers can use these models to predict the future labor market that can accommodate elderly in Thailand. Policymakers can apply our models (lasso logistic regression and random forest) to a well-collected sample of the population aged 50 to 60 to predict whether they are likely to continue work after sixty. The policymaker can make a long-term plan for the elderly's labor market based on their characteristics and backgrounds. Since 2017, Thailand Revenue Department encourages companies to hire elderly as it will qualify those companies for income tax exemption (\citealt{revenue_department_news_revenue_2017}). Several government sectors, Department of Employment, and Department of Older Persons, for instance, have introduced several measures to accommodate elderly to the labor market including upskilling and reskilling elderly with vocational training and information technology skills (\citealt{Department_of_Employment_2018}; \citealt{Department_of_Older_Persons_2021}). 
By adapting our model to each economic sector, we can help both the public and private sectors narrow down which skills should be highlighted. This will enable elderly to have the necessary skills and to adapt to the current labor market. Further, for elderly who are in the goods and services sectors, there should be an avenue specified only for elderly to minimize the competition which can ease the stress among the elderly in the economic sector. Moreover, by following our idea, a new model can be re-constructed each time a new data set is available. Doing so, we will have the most up-to-date model that accurately reflexes elderly's desire to work at a particular time.

An important step for future research on elderly's desire to work is to improve measures of work type. This study was limited by the data set utilized as National Statistical Office of Thailand does not have details of the work type. As mentioned, elderly who do not retire does not mean that they desire to work full-time, some elderly engage in bridge employment (\citealt{beehr_working_2015}; \citealt{quinn_work_2010}) while other elderly engage in unretirement (\citealt{maestas_back_2010}). Further, this study can be strengthened if other variables were incorporated in the survey such as the last job that elderly hold, years of work, desire jobs (if the answer is “desire to work”), for instance.   

\section*{Acknowledgements}

We are grateful for the dataset from the National Statistical Office of Thailand.

\section*{Declarations}

\begin{flushleft}
\textbf{Funding:} This work was financially supported by Office of the Permanent Secretary, Ministry of Higher Education, Science, Research and Innovation (Grant number: RGNS 63-217).

\textbf{Conflicts of interest/Competing interests:} No potential conflict of interest was reported by the authors.

\textbf{Availability of data and material:} This dataset is under the license of National Statistical Office of Thailand.

\textbf{Code availability:} Available upon request.

\textbf{Author's contributions:} Both authors have equal authorship of this article.

\textbf{IRB approval:} This research has IRB approvals, the certification ID is SWUEC/E-346/2564 and KMUTT-IRB-COE-2021-020.
\end{flushleft}

\end{document}